\documentclass[aps,showpacs,nofootinbib]{revtex4}

\usepackage{graphicx}

\newcommand{\be}{\begin{equation}}
\newcommand{\ee}{\end{equation}}
\newcommand{\ben}{\begin{eqnarray}}
\newcommand{\een}{\end{eqnarray}}

\newcommand{\la}{{\lambda}}

\newcommand{\cO}{{\cal O}}

\newcommand{\p}{\partial}
\newcommand{\na}{\nabla}

\newcommand{\tchi}{\tilde \chi}
\newcommand{\tim}{{\tilde \mu}}
\newcommand{\tom}{{\tilde \omega}}

\newcommand{\tV}{{\tilde V}}

\newcommand{\Dsl}{{\slash \negthinspace \negthinspace \negthinspace \negthinspace  D}}

\newcommand{\talpha}{{\tilde \alpha}}
\newcommand{\tbeta}{{\tilde \beta}}

\newcommand{\ep}{\epsilon}

\newcommand{\ga}{\gamma}


\begin{document}

\title{Decay of Dirac Massive Hair in the Background of Spherical Black Hole}

\author{ Rafa{\l} Moderski}
\address{
Nicolaus Copernicus Astronomical Center \protect \\
Polish Academy of Sciences \protect \\
00-716 Warsaw, Bartycka 18, Poland \protect \\
moderski@camk.edu.pl }

\author{Marek Rogatko}
\affiliation{Institute of Physics \protect \\
Maria Curie-Sklodowska University \protect \\
20-031 Lublin, pl.~Marii Curie-Sklodowskiej 1, Poland \protect \\
rogat@tytan.umcs.lublin.pl \protect \\
rogat@kft.umcs.lublin.pl}

\date{\today}

\begin{abstract}
The intermediate and late-time behaviour of massive Dirac hair in the
static spherically general black hole spacetime was studied. It was
revealed that the intermediate asymptotic pattern of decay of massive
Dirac spinor hair is dependent on the mass of the field under
consideration as well as the multiple number of the wave mode. The
long-lived oscillatory tail observed at timelike infinity in the
considered background decays slowly as $t^{- {5/6}}$.
    
\end{abstract}

\maketitle

\section{Introduction}
Late-time behaviour of various fields in the spacetime of a collapsing
body plays an important role in black hole physics.  It happens that
regardless of details of the collapse or the structure and properties
of the collapsing body the resultant black hole can be described only
by few parameters such as mass, charge and angular momentum, {\it
  black holes have no hair}.  On its own, it is interesting to
investigate how these hair loss proceeds dynamically.\par

\noindent
The research in this direction have quite long history.  Price in
\cite{pri72} showed that the late-time behavior is dominated by the
factor $t^{-(2l + 3)}$, for each multipole moment $l$.  It was
revealed in Ref.~\cite{gun94} that the decay-rate along null infinity
and along the future event horizon was governed by the power laws
$u^{-(l + 2)}$ and $v^{-(l + 3)}$, where $u$ and $v$ were the outgoing
Eddington-Finkelstein (ED) and ingoing ED coordinates.  On the other
hand, Ref.~\cite{bic72} was devoted to the scalar perturbations on
Reissner-Nordtr\"om (RN) background for the case when $\mid Q \mid <
M$ has the following dependence on time $t^{-(2l + 2)}$, while for
$\mid Q \mid = M$ the late-time behavior at fixed $r$ is governed by
$t^{-(l + 2)}$.\par

\noindent
A charged hair turned out to decay slower than a neutral one
\cite{pir1}-\cite{pir3}.  The late-time tails in gravitational
collapse of a self-interacting (SI) fields in the background of
Schwarzschild solution was reported by Burko~\cite{bur97} and in RN
solution at intermediate late-time was considered in
Ref.~\cite{hod98}.  At intermediate late-time for small mass $m$ the
decay was dominated by the oscillatory inverse power tails $t^{-(l
+3/2)} \sin (m t)$.  This analytic prediction was verified at
intermediate times, where $mM \ll mt \ll 1/(mM)^2$.  It was proved
analytically \cite{ja} that for a nearly extreme RN spacetime the
inverse power law behavior of the dominant asymptotic tail is of the
form $t^{-5/6} \sin (m t)$.  The asymptotic tail behaviour of SI scalar
field was also studied in Schwarzschild spacetime \cite{ja1}. The
oscillatory tail of scalar field has the decay rate of $t^{-5/6}$ at
asymptotically late time.  The power-law tails in the evolution of a
charged massless scalar field around a fixed background of dilaton
black hole was studied in Ref.~\cite{mod01a}, while the case of a
self-interacting scalar field was elaborated in \cite{mod01b}.  The
analytical proof of the above mentioned behaviour of massive scalar
hair in the background of dilaton black hole in the theory with
arbitrary coupling constant between $U(1)$ gauge field and dilaton
field was given in Ref.~\cite{rog07}.\par

\noindent
On its own, the late-time behaviour of massive Dirac fields were
studied in the spacetime of Schwarzschild black hole \cite{jin04}.  It
was found that the asymptotic behaviour of these fields is dominated
by a decaying tail without any oscillation.  The dumping exponent was
independent on the multiple number of the wave mode and on the mass of
the Dirac field.  It also happened that the decay of the massive Dirac
fields was slower comparing the decay of massive scalar field.  The
above analysis was supplemented by the studies of charged massive
Dirac fields in the spacetime of RN black hole \cite{jin05}.  The case
of the decay of the fields in the stationary axisymmetric black hole
background was studied numerically in Ref.~\cite{bur04} and it was
found that in the case of Kerr black hole that the oscillatory
inverse-power law of the dominant asymptotic tail behaviour is
approximately depicted by the relation $t^{-5/6}\sin(mt)$.  In
Ref.~\cite{xhe06} both the intermediate late-time tail and the
asymptotic behaviour of the charged massive Dirac fields in the
background of Kerr-Newmann black hole was investigated.  It was
demonstrated that the intermediate late-time behaviour of the fields
under consideration is dominated by an inverse power-law decaying tail
without any oscillation.\par

\noindent
The late-time behaviour of massive vector field obeying the Proca
equation of motion in the background of Schwarzschild black hole was
studied in \cite{kon06}.  It was revealed that at intermediate late
times, three functions characterizing the field have different decay
law depending on the multiple number $l$.  On the contrary, the
late-time behaviour is independent on $l$, i.e., the late-time decay
law is proportional to $t^{-5/6}\sin(mt)$.  In Ref.~\cite{gibrog08} the
analytical studies concerning the intermediate and late-time decay
pattern of massive Dirac hair on the dilaton black hole were
conducted.  Dilaton black hole constitutes a static spherically
symmetric solution of the theory being the low-energy limit of the
string theory with arbitrary coupling constant $\alpha$.\par

\noindent
For the case of $n$-dimensional static black holes, the {\it no-hair}
theorem is quite well established \cite{unn}.  The mechanism of
decaying black hole hair in higher dimensional static black hole case
concerning the evolution of massless scalar field in the
$n$-dimensional Schwarzshild spacetime was determined in
Ref.~\cite{car03}.  The late-time tails of massive scalar fields in the
spacetime of $n$-dimensional static charged black hole was elaborated
in Ref.~\cite{mod05} and it was found that the intermediate asymptotic
behaviour of massive scalar field had the form $t^{-(l + n/2 -
1/2)}$.  This pattern of decay was checked numerically for $n = 5, 6$.
Quasi-normal modes for massless Dirac field in the background of
$n$-dimensional Schwarzschild black hole were obtained in
\cite{cho07a}, while in Ref.~\cite{cho07b} studies of massless fermion
excitations on a tensional three-brane embedded in six-dimensional
spacetime were provided.  On the other hand, the intermediate and
late-time decay of massive scalar hair on static brane black holes was
elaborated in \cite{rog07br}, while the decay of massive Dirac hair in the spacetime of black hole in question
was studied in \cite{gibrog08br}.\par

\noindent
At the beginning of our universe several phase transitions leading to
the topological defect formations might happened \cite{vil}.  The
interactions of topological defects such as global monopoles or cosmic
strings with compact objects like black holes attract much attention
and their physical characteristics are widely studied in literature.
For instance, the black hole global monopole system has an unusual
topological property of possessing a solid deficit angle \cite{mon}.
The decay of massive scalar hair in the background of a Schwarzschild
black hole with global monopole was considered in Ref.~\cite{hyu02}.
It happened that the topological defects makes the massive scalar
field hair decay faster in the intermediate regime comparing to the
decay of such hair on the black holes without defects.  On the other
hand, the late-time behaviour was unaffected by the presence of global
monopole.  The Schwarzschild black hole global monopole system coupled
to scalar fields was elaborated in Ref.~\cite{che05}.\par

\noindent
Cosmic strings also acquire much interests as well as the cosmic
string black hole systems. The metric of such a system assuming the
distributional mass source was derived in \cite{ary86} (the so-called
{\it thin string limit}).  Then it was revealed that it constituted
the limit of much more realistic situation when black hole was pierced
by a Nielsen-Olesen vortex \cite{greg}.  Especially interesting
behaviour was found in the case of extremal black hole.  Namely, for
some range of black hole parameters it was shown that extremal black
hole expelled the vortex (the so called {\it Meissner effect}).
It turned our that in the case of extremal black holes
in dilaton gravity one has always expulsion of the Higgs fields from
their interiors \cite{mod99}.  The intermediate and late-time decay
patterns of massive Dirac hair in the spacetime of black hole and
topological defects were elaborated in Ref.~\cite{rog08t}.  The system of
black hole with a global monopole and dilaton black hole pierced by a
cosmic string was considered.  Among all it was revealed that the
late-time asymptotic decay of massive Dirac hair in the systems in
question were proportional to $t^{- 5/6}$ and topological defects did
not affect this relation.\par

\noindent
In our paper we shall discuss the intermediate and
late-time behaviour of the massive Dirac spinor field in the spacetime
of static spherically symmetric general kind of black hole.  We
revealed that the intermediate asymptotic behaviour is not the final
pattern of the decay of the massive spinor hair.  In Sec.~II we present
and summary our analytic arguments concerning the decay of massive
spinor Dirac hair in the background of the black hole under
consideration.  In Sec.~III we conclude that the long-lived
oscillatory tail is generally observed at timelike infinity in static
spherically symmetric black hole spacetime which decay pattern is
proportional to $t^{- {5 \over 6}}$.

\section{ Massive Dirac hair on spherically symmetric black hole}

\subsection{Properties of the Dirac Equation}
The mass Dirac equation in a background of spherically symmetric black
hole is given by
\be \bigg( \ga^{\mu} \na_{\mu}\psi - m \bigg) \psi = 0, \ee
where $\na_{\mu}$ is the covariant derivative $\na_{\mu} = \p_{\mu} +
{1\over 4} \omega_{\mu}^{ab} \ga_{a} \ga_{b}$, $\mu$ and $a$ are
tangent and spacetime indices.  There are related by $e_{\mu}^{a}$, a
basis of orthonormal one-forms.  The quantity $\omega_{\mu}^{ab}
\equiv \omega^{ab}$ are the associated connection one-forms satisfying
$de^{a} + \omega_{b}{}{}^{a} \wedge e^{b} = 0$, while the gamma
matrices fulfill the relation $\{ \ga^{a}, \ga^{b} \} = 2 \eta^{ab}$.
Now we shall recall some basic properties of the Dirac operator $\Dsl
= \gamma ^\mu \nabla _\mu $ on an $n$-dimensional manifold (see e.g.,
\cite{gibrog08}).  In what follows we assume that the metric of the
underlying spacetime may be rewritten as a product of the form
\be g_{\mu \nu} d x ^\mu d x ^\nu = g_{ab} (x) dx^a dx ^b + g_{mn}(y)
dy^m dy ^n.  \ee
The above metric decomposition will be subject to the direct sum of
the Dirac operator, namely one obtains
\be \Dsl = \Dsl_x + \Dsl_y.  \ee
If we define a Weyl conformal rescaling as
\be g_{\mu \nu} = \Omega ^2 {\tilde g} _{\mu \nu}, \ee
it consequently provides that one gets
\be \Dsl \psi =\Omega ^{- {1 \over 2} (n+1)} { \tilde {\Dsl} } {\tilde
\psi}\,, \qquad \psi = \Omega ^ {- {1 \over 2} (n-1) }\tilde \psi.
\ee
Having in mind that spherically symmetric form of the line element
provides also conformal flatness for a static metric, one obtains
\be ds^2 = - A^2 dt^2 + B^2 dr ^2 + C^2 d \Sigma ^2 _{n-2} \,,  \ee
where $A=A(r)$, $B=B(r)$, $C=C(r)$ are functions only of the radial
variable $r$, and the {\it transverse} metric $d \Sigma ^2 _{n-2} $ is
independent on $t$ and on $r$.

\noindent
Suppose then, that $\Psi$ is a spinor eigenfunction on the
$(n-2)$-dimensional {\it transverse} manifold $\Sigma$.  It leads to
the following:
\be \Dsl_\Sigma \Psi = \lambda \Psi.  \ee
In case of $(n - 2)$-dimensional sphere the eigenvalues for spinor $\Psi$,
where found in Ref.\cite{cam96}. They imply
\be
\la^2 = \bigg( l + { n - 2 \over 2} \bigg)^2,
\ee
where $l = 0, 1, \dots$
Using the properties given above one may also assume that the
following is satisfied:
\be \Dsl \psi = m \psi.  \ee
It enables us to set the form of the spinor $\psi$, i.e.,
\be \psi = {1 \over A^{1 \over 2} } { 1 \over C^{(n-2) \over 2 }} \chi
\otimes \Psi.  \ee
Next, the explicit calculations reveal that we arrive at the relation
\be (\gamma ^0 \partial _t + \gamma ^1 \partial _y) \chi = A (m-
    {\lambda \over C} ) \chi \,, \ee
where we have introduced the {\it radial optical distance} (i.e., the
Regge-Wheeler radial coordinate) $dy = {B / A} dr$ and $\gamma^0,
\gamma^1$ satisfy the Clifford algebra in two spacetime
dimensions.\par

\noindent
We remark that an identical result may be obtained if a Yang-Mills
gauge field $A_{\mu}$ is present on the {\it transverse} manifold
$\Sigma$.  The only difference is that
\be \Dsl_{\Sigma, A_{\mu}} \Psi = \lambda \Psi \,, \ee
where $ \Dsl_{\Sigma, A_{\mu}} $ is the Dirac operator twisted by the
the connection $A_{\mu}$.  Assuming that $\psi \propto e^{-i\omega t}$
one obtains the second order equation of the form
\be {d^2 \chi \over d y^2 }+ \omega ^2 \chi = A^2 (m-{\lambda \over C}
)^2 \chi.
\label{second}
\ee

\subsection{The Background of Spherically Symmetric Static Black Hole}

As was shown in Refs.~\cite{hod98,lea86} the spectral decomposition
method will be fruitful for the analysis of the evolution of massive
Dirac hair in the spacetime of static spherically symmetric black
hole.  The asymptotic tail is connected with the existence of a branch
cut situated along the interval $-m \le \omega \le m$.  An oscillatory
inverse power-law behaviour of the massive spinor field arises from
the integral of Green function $\tilde G(y, y';\omega)$ around branch
cut.  The time evolution of the field may be written in the following
form:
\be \chi(y, t) = \int dy' \bigg[ G(y, y';t) \psi_{t}(y', 0) + G_{t}(y,
  y';t) \psi(y', 0) \bigg], \ee
for $t > 0$, where the Green's function $ G(y, y';t)$ is given by the
relation
\be \bigg[ {\p^2 \over \p t^2} - {\p^2 \over \p y^2 } + V \bigg] G(y,
y';t) = \delta(t) \delta(y - y').
\label{green}
\ee
Next, our main task will be to find the black hole Green function.  In
the first step we reduce equation (\ref{green}) to an ordinary
differential equation.  To do it one can use the Fourier transform
\cite{lea86} $\tilde G(y, y';\omega) = \int_{0^{-}}^{\infty} dt~ G(y,
y';t) e^{i \omega t}$.  This Fourier's transform is well defined for
$Im~ \omega \ge 0$, while the corresponding inverse transform yields
\be G(y, y';t) = {1 \over 2 \pi} \int_{- \infty + i \ep}^{\infty + i
  \ep} d \omega~ \tilde G(y, y';\omega) e^{- i \omega t}, \ee
for some positive number $\ep$.  The Fourier's component of the
Green's function $\tilde G(y, y';\omega)$ can be written in terms of
two linearly independent solutions for homogeneous equation as
\be \bigg( {d^2 \over dy^2} + \omega^2 - \tV \bigg) \chi_{i} = 0,
\qquad i = 1, 2 \,,
\label{wav}
\ee
where $\tV = A^{2} \bigg( m - {\la \over C} \bigg)^2$.

\noindent
The boundary conditions for $\psi_{i}$ are described by purely ingoing
waves crossing the outer horizon $H_{+}$ of the $n$-dimensional static
charged black hole $\psi_{1} \simeq e^{- i \omega y}$ as $y
\rightarrow - \infty$ while $\psi_{2}$ should be damped exponentially
at $i_{+}$, namely $\psi_{2} \simeq e^{- \sqrt{m^2 - \omega^2}y}$ at
$y \rightarrow \infty$.

In order to proceed further, we change the variables
\be \chi_{i} = {\xi \over A^{1/2} ~B^{- 1/2}} \,, \ee
where $i = 1,2$. 
\par
Although the formalism presented above is true for an arbitrary $n$-dimensional
spacetime we shall consider general form of four-dimensional spherically symmetric static
black hole. Black hole under consideration will be characterized by gravitational mass
$M$ and some other parameters $M',~Q,~Q'$ of the background field in addition to
the gravitational mass.\\
We shall consider wave modes only in a far region from the black hole,
as a generic behaviours observed in any black hole background. For
that purpose, let us assume that $r/M \gg 1$, where $M$ is the
gravitational mass of background field.  Having all these in mind, the
expansion of the metric coefficients $A(r)$ and $B(r)$ as a power
series in $M/r$ give us the following:
\be A(r) = 1 - {2 M \over r} + {Q^2 \over r^2} + \cO \bigg( {1 \over
  r^3} \bigg), \ee
while for the next metric function one has the relation
\be B(r) = 1 + {2 M' \over r} + {Q'^2 \over r^2} + \cO \bigg( {1 \over
  r^3} \bigg), \ee
where by $M'$ and $Q$, $Q'$ we have denoted other parameters
characterizing the background fields.
\par
Let us assume that the observer and the initial data are situated far
away from the considered black hole.  We expand Eq.~(\ref{wav}) as
power series of $ M/r$ neglecting terms of order $\cO (\omega/r)^2)$
and higher.  It leads us to the following expression:
\ben {d^2 \over dr^2} \xi + \bigg( \omega^2 - m^2 \bigg) \xi &+&
\bigg[ {2 \talpha (\omega^2 - m^2 ) + \alpha_{1} \over r} \\ \nonumber
  &-& { \omega^2 ( \talpha^2 + \tbeta^2 ) + \beta_{1} - m^2~\talpha^2
    - 2 \talpha \alpha_{1} - 2 \tbeta m^2 \over r^2} \bigg] \xi = 0
\,, \een
where 
\ben \alpha_{1} &=& 2m (\la + Mm), \qquad \beta_{1} = \la^2 + 4 Mm \la
+ Q^2 m^2, \\ \nonumber \talpha &=& M + M', \qquad \tbeta = {Q^2 -
  Q'^2 \over 2} + {M'^2 - M^2 \over 2} + MM'.  \een
It turned out that the above equation can be solved in terms of
Whittaker's functions, namely the two basic solutions are needed to
construct the Green function, with the condition that $\mid \omega
\mid \ge m$.  Consequently, it implies the result as follows:
\be \tchi_{1} = M_{\kappa, \tim}(2 \tom r), \qquad \tchi_{2} =
W_{\kappa, \tim}(2 \tom r) \,, \ee
where we have denoted
\ben \tim &=& \sqrt{ 1/4 + \omega^2 ( \talpha^2 + \tbeta^2 ) +
  \beta_{1} - m^2~\talpha^2 - 2 \talpha \alpha_{1} - 2 \tbeta m^2 },
\\ \nonumber \kappa &=& {2 \talpha (\omega^2 - m^2 ) + \alpha_{1}
  \over 2 \tom}, \qquad \tom^2 = m^2 - \omega^2.  \een
By virtue of the above relations the spectral Green function has the
form
\ben G_{c}(x,y;t) &=& {1 \over 2 \pi} \int_{-m}^{m}dw \bigg[
  {\tchi_{1}(x, \tom e^{\pi i})~\tchi_{2}(y,\tom e^{\pi i}) \over
    W(\tom e^{\pi i})} - {\tchi_{1}(x, \tom )~\tchi_{2}(y,\tom ) \over
    W(\tom )} \bigg] ~e^{-i w t} \\ \nonumber &=& {1 \over 2 \pi}
\int_{-m}^{m} dw f(\tom)~e^{-i w t}, \een
where $W(\tom)$ is the Wronskian.

\noindent
First we discuss the intermediate asymptotic behaviour of the massive
scalar field, i.e., when the range of parameters are $M \ll r \ll t
\ll M/(m M)^2$.  The intermediate asymptotic contribution to the Green
function integral gives the frequency equal to $\tom = {\cO
  (\sqrt{m/t})}$.  It implies that $\kappa \ll 1$.  One should have in
mind that $\kappa$ stems from the $1/r$ term in the massive scalar
field equation of motion.  Thus it depicts the effect of
backscattering off the spacetime curvature and in the case under
consideration the backscattering is negligible.  In the case of
intermediate asymptotic behaviour one finally gets
\be f(\tom) = {2^{2 \tim -1} \Gamma(-2\tim)~\Gamma({1 \over 2} + \tim)
  \over \tim \Gamma(2 \tim)~\Gamma({1 \over 2} - \tim)} \bigg[ 1 +
  e^{(2 \tim + 1) \pi i} \bigg] (r r')^{{1 \over 2} + \tim} \tom^{2
  \tim}, \ee
where we have used the fact that $\tom r \ll 1$ and the form of
$f(\tom)$ can be approximated by means of the fact that $M(a, b, z) =
1$ as $z$ tends to zero.\par

\begin{figure}
\begin{center}
  \includegraphics[width=0.85\textwidth]{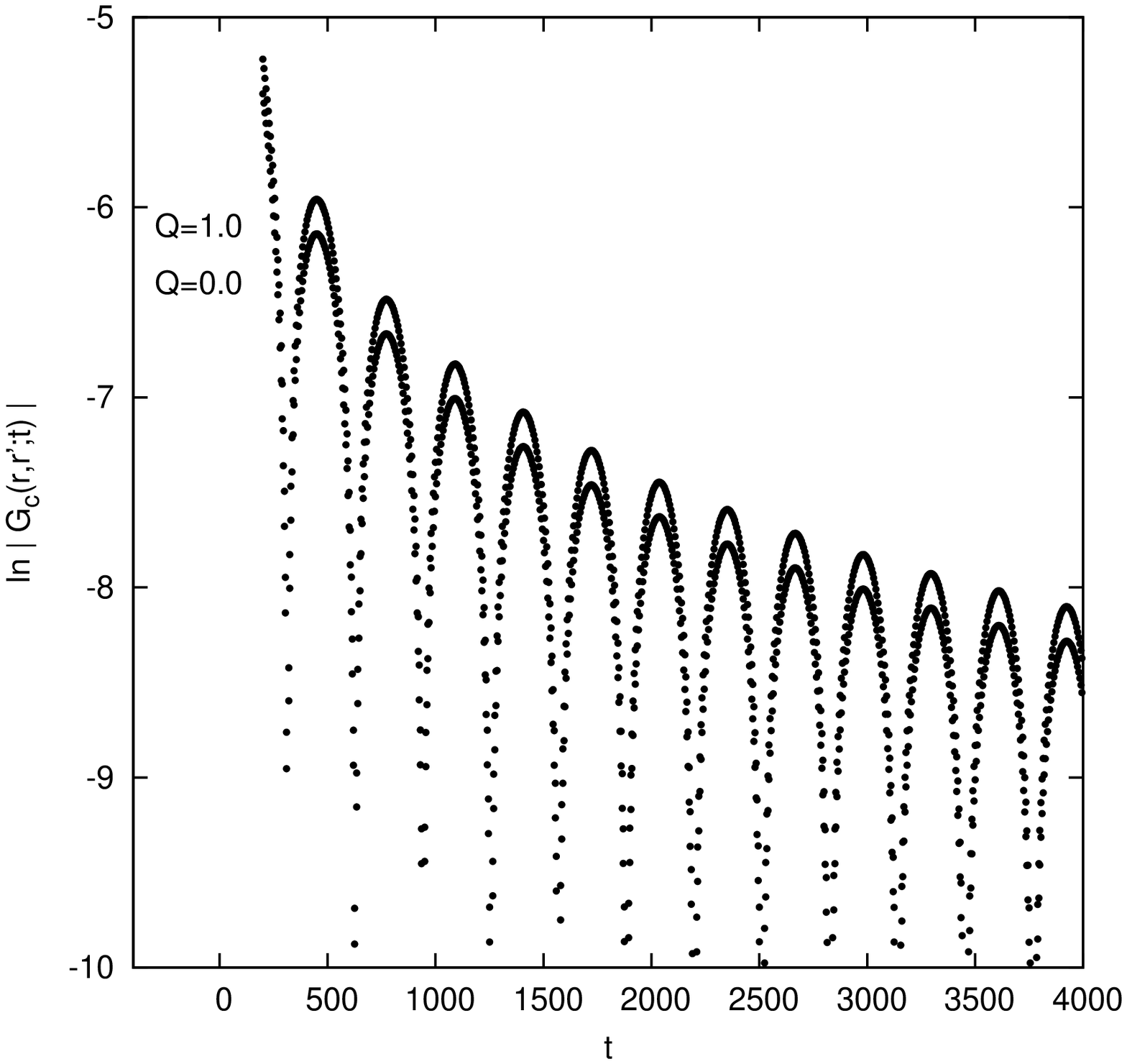}
\end{center}
\caption{Amplitude of the Green's function vs. time for different $Q$.
  Parameters: $M=1$, $M'=1$, $Q'=0$, $rr'=1000$, $\lambda =
  0.01$, $m=0.01$}
\label{fig:fig1}
\end{figure}

\begin{figure}
\begin{center}
  \includegraphics[width=0.85\textwidth]{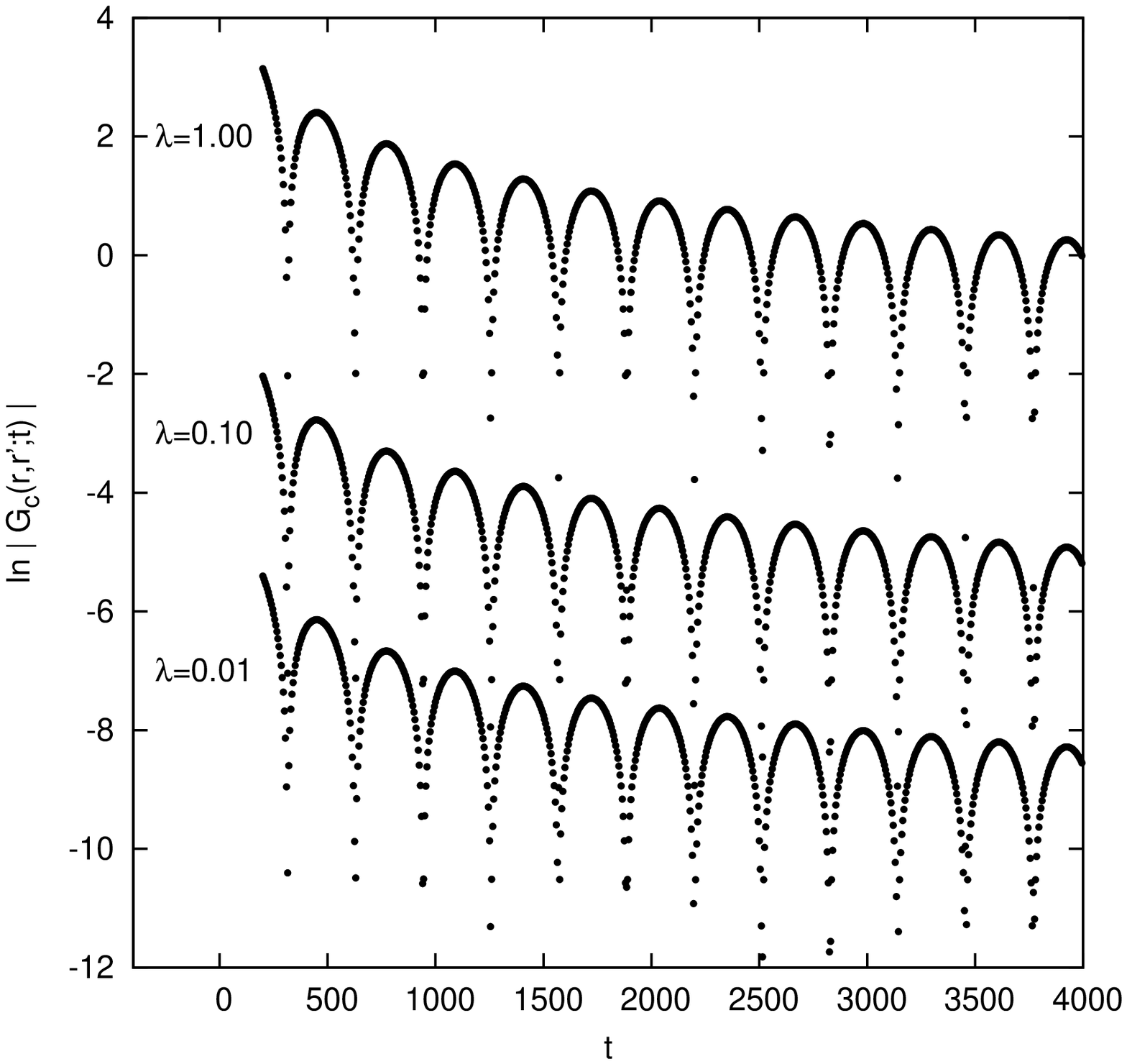}
\end{center}
\caption{Amplitude of the Green's function vs. time for different
  $\lambda$.  Parameters: $M=1$, $M'=1$, $Q=0$, $Q'=0$, $rr'=1000$,
  $m=0.01$.}
\label{fig:fig2}
\end{figure}

\begin{figure}
\begin{center}
  \includegraphics[width=0.85\textwidth]{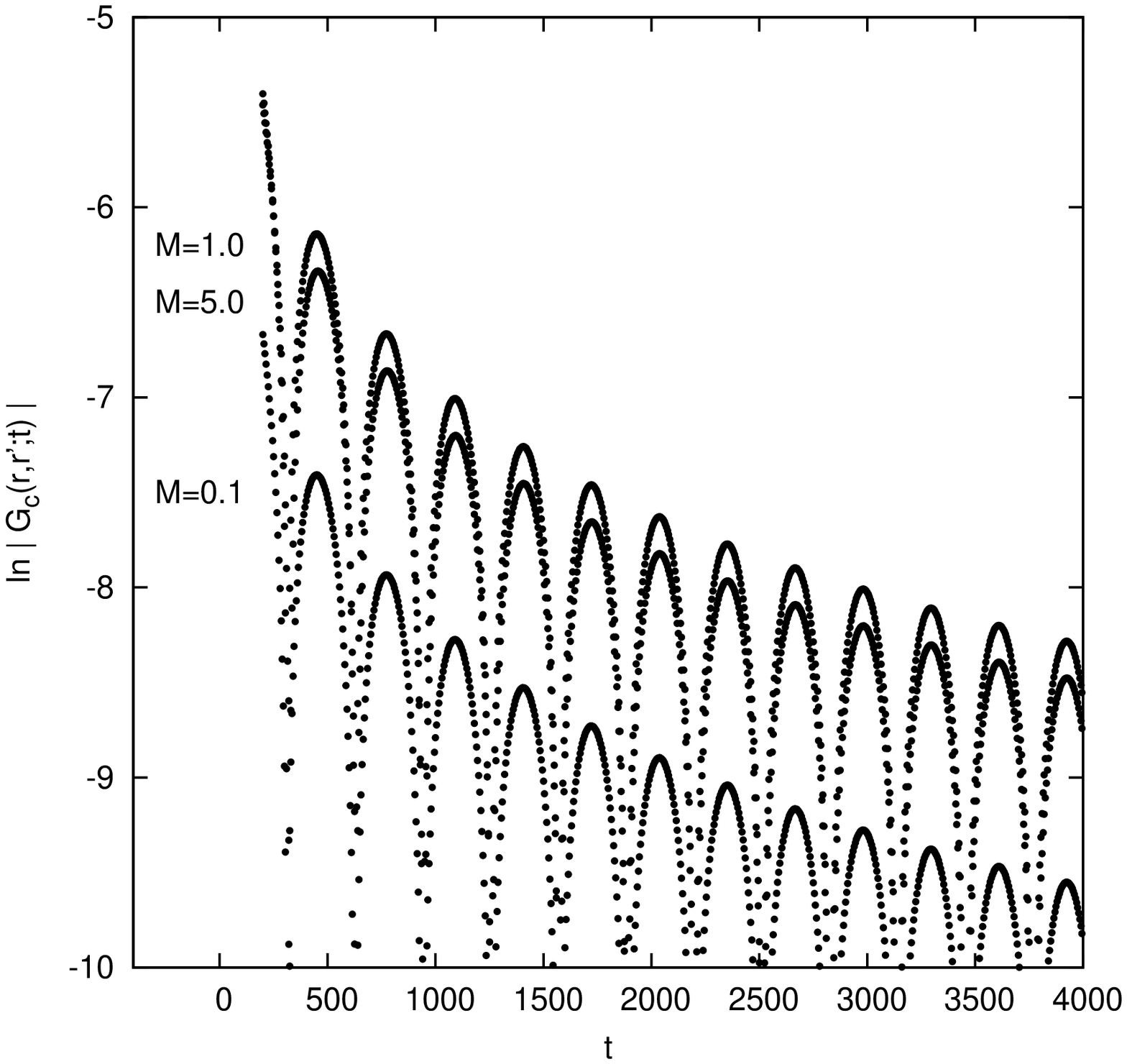}
\end{center}
\caption{Amplitude of the Green's function vs. time for different $M$.
  Parameters: $M'=1$, $Q=0$, $Q'=0$, $rr'=1000$, $\lambda=0.01$,
  $m=0.01$.}
\label{fig:fig3}
\end{figure}

\begin{figure}
\begin{center}
  \includegraphics[width=0.85\textwidth]{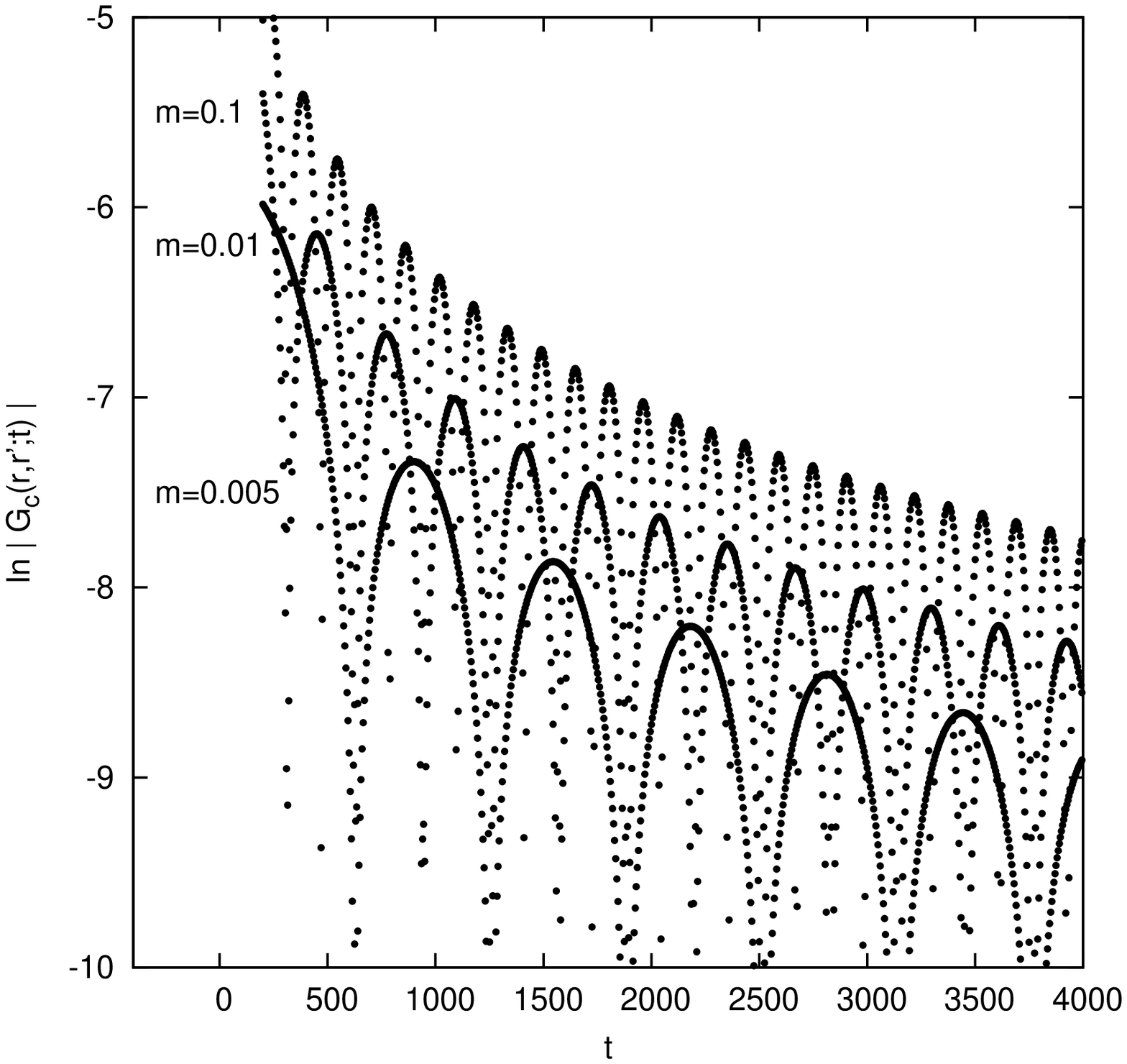}
\end{center}
\caption{Amplitude of the Green's function vs. time for different $m$.
  Parameters: $M=1$, $M'=1$, $Q=0$, $Q'=0$, $rr'=1000$,
  $\lambda=0.01$.}
\label{fig:fig4}
\end{figure}

\begin{figure}
\begin{center}
  \includegraphics[width=0.85\textwidth]{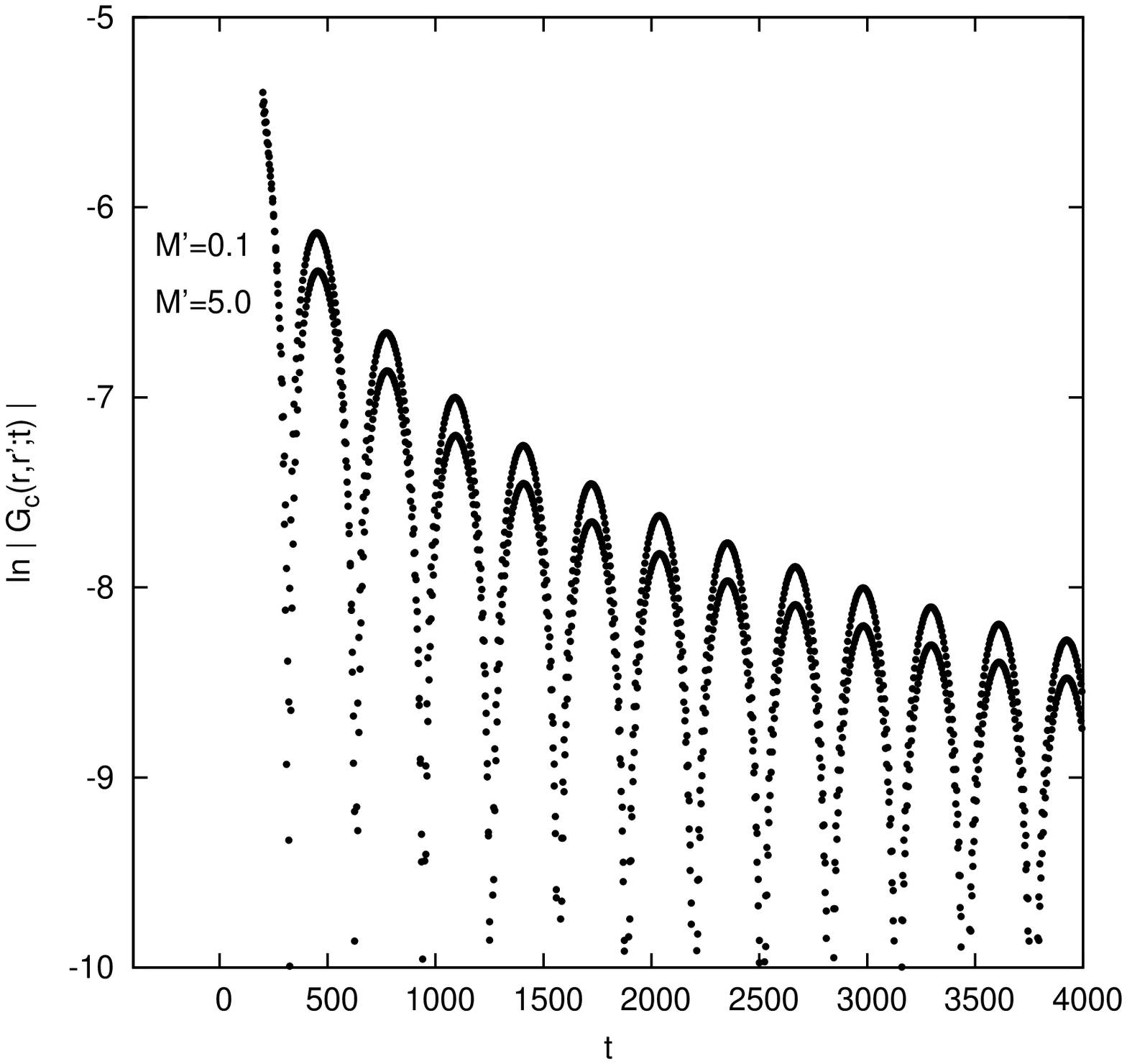}
\end{center}
\caption{Amplitude of the Green's function vs. time for different $M'$.
  Parameters: $M=1$, $Q=0$, $Q'=0$, $rr'=1000$, $\lambda=0.01$,
  $m=0.01$.}
\label{fig:fig5}
\end{figure}

\noindent
However, the resulting Green function can not be calculated
analytically because of the fact that the parameter $\tim$ depends on
$\omega$.  We calculate it numerically and results are provided in
Figs~\ref{fig:fig1}-\ref{fig:fig5}.  In Fig.~1 we present the graph of
the logarithm of intermediate Green function $\ln \mid G_{c}(r, r';t)
\mid$ versus time $t$ for different mass parameter $M = 0.1, M = 1.0 $
and $M = 5.0$. In Fig.~2 we depict the dependence of the function in
question on $M' = 0.1, M' = 5.0$, while in Fig.~3 we get the relation
among intermediate Green function and $Q$ (for $Q = 0.0$ and $Q =
1.0$).  Fig.~4 is connected with the dependence of the intermediate
Green function on mass of the Dirac hair.  We plotted the graph for $m
= 0.005,~m = 0.01,~m = 0.02$.  We finished with the Fig.~5, where the
dependence on the multiple number of the wave mode was revealed for
$\la = 0.01, \la = 0.1$ and $\la = 1.0$.  Summing it all up one can
remark that the intermediate asymptotic Green function depends on
the parameters $M$, $M'$, $Q$ as well as the mass of the Dirac spinor
field and the multiple number of the wave mode.\par

\noindent
The different pattern of decay is expected when $\kappa \gg 1$, for
the late-time behaviour, when the backscattering off the curvature is
important.  In this case we use the limit
\be M_{\kappa, \tim}(2 \tom r) \approx \Gamma (1 + 2 \tim)~(2 \tom
r)^{1 \over 2} \kappa^{- \tim}~J_{2 \tim} (\sqrt{8 \kappa \tom r}).
\ee
Consequently, $f(\tom)$ yields
\ben \label{fer} f(\tom) &=& {\Gamma(1 + 2\tim)~\Gamma(1 - 2\tim)
  \over 2 \tim}~(r r')^{1 \over 2} \bigg[ J_{2 \tim} (\sqrt{8 \kappa
    \tom r})~J_{- 2 \tim} (\sqrt{8 \kappa \tom r'}) - I_{2 \tim}
  (\sqrt{8 \kappa \tom r})~I_{- 2 \tim} (\sqrt{8 \kappa \tom r'})
  \bigg] \\ \nonumber &+& {(\Gamma(1 + 2\tim))^2~\Gamma(-
  2\tim)~\Gamma( {1 \over 2} + \tim - \kappa) \over 2 \tim ~\Gamma(2
  \tim)~\Gamma({1 \over 2} - \tim - \kappa) }~(r r')^{1 \over 2}
~\kappa^{- 2 \tim} \bigg[ J_{2 \tim} (\sqrt{8 \kappa \tom r})~J_{2
    \tim} (\sqrt{8 \kappa \tom r'}) \\ \nonumber &+& e^{(2 \tim + 1)}
  I_{2 \tim} (\sqrt{8 \kappa \tom r})~I_{2 \tim} (\sqrt{8 \kappa \tom
    r'}) \bigg].  \een
It can be noticed that the first part of the above Eq.~(\ref{fer}) the
late time tail is proportional to $t^{-1}$.  Just we calculate the
second term of the right-hand side of Eq.~(\ref{fer}).  For the case
when $\kappa \gg 1$ it can be brought to the form written as
\be G_{c~(2)}(r, r';t) = {M \over 2 \pi} \int_{-m}^{m}~dw~e^{i (2 \pi
  \kappa - wt)}~e^{i \phi}, \ee
where we have defined
\be e^{i \varphi} = { 1 + (-1)^{2 \tim} e^{- 2 \pi i \kappa} \over 1 +
  (-1)^{2 \tim} e^{2 \pi i \kappa}}, \ee
while $M$ provides the relation as follows:
\be M = {(\Gamma(1 + 2\tim))^2~\Gamma(- 2\tim) \over 2 \tim ~\Gamma(2
  \tim) }~(r r')^{1 \over 2} \bigg[ J_{2 \tim} (\sqrt{8 \kappa \tom
    r})~J_{2 \tim} (\sqrt{8 \kappa \tom r'}) + I_{2 \tim} (\sqrt{8
    \kappa \tom r})~I_{2 \tim} (\sqrt{8 \kappa \tom r'}) \bigg].  \ee
At very late time both terms $e^{i w t}$ and $e^{2 \pi \kappa}$ are
rapidly oscillating.  It means that the scalar waves are mixed states
consisting of the states with multipole phases backscattered by
spacetime curvature.  Most of them cancel with each others which have
the inverse phase.  In such a case, one can find the value of $G_{c
  (2)}$ by means of the saddle point method.  It could be found that
the value $2 \pi \kappa - wt$ is stationary at the value of $w$ equal
to the following:
\be a_{0} = \bigg[ { \pi~(2 \talpha (\omega^2 - m^2 ) + \alpha_{1})
    \over 2 \sqrt{2} m} \bigg]^{1 \over 3}, \ee
Then approximating integration by the contribution from the very close
nearby of $a_{0}$ is given by
\be F = {i m^{2/3} \sqrt{2} \over 3 \sqrt{\alpha_{1}}}
(\pi~\alpha_{1})^{5 \over 6} (mt)^{-{ 5 \over 6}}~e^{i
  \varphi(a_{0})}.  \ee
In comparison to the late-time behaviour of the second term in
Eq.~(\ref{fer}), the first term can be neglected.  The dominant role
plays the behaviour of the second term, i.e., the late-time behaviour
is proportional to ${- {5 \over 6}}$.  Thus, the asymptotic late-time
behaviour of the Green's function can be written in the form
\be G_{c}(r,r;t) = {2 \sqrt{2}~ m^{2/3} \over \sqrt{3}} \pi^{5 \over
  6}~\alpha_{1}^{1 \over 3}~ (mt)^{-{ 5 \over 6}}~\sin(mt)~\tchi(r,
m)~\tchi(r', m).  \ee
The above relation provides the main conclusion of our investigations,
i.e., analytical proof that the late-time asymptotic decay pattern of
massive Dirac field is unaffected by black hole characteristics and
has the form of power law pattern proportional to $t^{-5/6}$.

\section{Conclusions}
Dirac fields were intensively investigated in various contexts in spacetimes of black holes. Among all, problems 
concerning
quasi-normal modes, high overtones were studied
(see e.g., the latest works concerning this widely elaborated field of research \cite{dup}).
In our paper we restrict our attention to the problem of the intermediate and late-time decay pattern of 
massive Dirac hair in the background of static spherically black hole. This problem is strictly
bounded with the {\it no-hair} theorem for the objects in question.
\par
It has been shown in previous papers \cite{jin04,jin05} that the oscillatory power-law decay
rate of massive Dirac hair which decay rate is $t^{-5/6}$ dominates at asymptotically late-time
in Schwarzschild and RN black hole spacetimes. In our considerations we envisaged the fact
that this behaviour was generic one observed in any static spherically symmetric background. We assume
that a massive Dirac field propagates in a static spherically symmetric spacetime with
asymptotically flat metric. The metric under consideration is described in terms of the ADM
mass $M$
and some other parameters $M',~Q,~Q'$ of the background fields in addition to the afore mentioned
gravitational mass. In Refs.\cite{jin04,jin05,xhe06} the complicated equation of motion for massive Dirac fields were
solved by means of the so-called Newman-Penrose formalism. In our research we introduce much easier method of 
solving Dirac equation.
The properties of the Dirac operator enable us to find the second order differential equation which in turn
can be applied to construct
the spectral Green function, crucial in our investigations.
One should remark that our method may be used in arbitrary spacetime dimension contrary to the method of solvig Dirac Eqs.
presented in the afore mentioned papers. However, we restrict our research to four-dimensional case.
\par
Unfortunately, the intermediate asymptotic behavior of massive spinor hair can not be calculated analytically.
Then, we performed numerical calculations and found
the graphs of the logarithm of the intermediate Green function versus time for
changes of the parameters like $M,~M',~Q$, mass of the Dirac field $m$
and the multiple number of the wave function.  We revealed the
dependence of the Green function in question on the above
parameters.
\par
However, it turned out this is not the final pattern of vanishing the
massive spinor hair.  The resonance backscattering off the spacetime
curvature reveals at very late times.  It was proved analytically
that the massive spinor decay in the spacetime under consideration is
proportional to $t^{- {5 \over 6}}$.  So we conclude that this
long-lived oscillating tail is generally observed at timelike infinity
in the black hole spacetimes.  The same conclusion concerning the
pattern of decay was achieved in Ref.~\cite{koy02}, where the massive
scalar hair decay was elaborated in static spherically symmetric black
hole spacetime.
\par
Our work comprises all the previously studied cases of static spherically
symmetric black hole spacetimes on which propagates massive Dirac field.
Finally, one should stress that we obtain the analytical proof that the late-time
decay pattern of massive Dirac spinor hair in the background of general static spherically
symmetric four-dimensional black hole it of the form $t^{- {5 \over 6}}$. It does not depend
on gravitational mass and additional parameters characterizing background fields.\\
It will be interesting to study the decay pattern of higher spin fields in the general static spherically 
background (the research in this direction was begun in Ref.\cite{kon06}) and to consider
$n$-dimensional black hole background. We hope to return to these problems elsewhere.

\begin{acknowledgments}
MR is grateful for hospitality of DAMTP, Center for Mathematical
Sciences, Cambridge, where the part of the research was begun.  This
work was partially financed by the Polish budget funds in 2008 year as
the research project.
\end{acknowledgments}



\pagebreak

\pagebreak

\pagebreak

\pagebreak

\pagebreak

\end{document}